\documentclass[twocolumn,english,prx,notitlepage,longbibliography,superscriptaddress]{revtex4-1}
\usepackage{mathptmx}
\usepackage[T1]{fontenc}
\usepackage[latin9]{inputenc}
\usepackage{amsmath}
\usepackage{graphicx}

\makeatletter
\usepackage{babel}
\usepackage{color}

\definecolor{green}{rgb}{0,0.7,0.3}

\usepackage{babel}

\makeatother

\usepackage{babel}
\begin{document}

\title{Neural-network states for the classical simulation of quantum computing }

\author{Bjarni J\'onsson }

\affiliation{Institute for Theoretical Physics, ETH Zurich, Wolfgang-Pauli-Str. 27,
8093 Zurich, Switzerland}

\author{Bela Bauer}

\affiliation{Station Q, Microsoft Corporation, Santa Barbara, California 93106,
USA}

\author{Giuseppe Carleo}

\affiliation{Center for Computational Quantum Physics, Flatiron Institute, 162
5th Avenue, New York, NY 10010, USA}
\begin{abstract}
Simulating quantum algorithms with classical resources generally requires
exponential resources. However, heuristic classical approaches are
often very efficient in approximately simulating special circuit structures,
for example with limited entanglement, or based on one-dimensional
geometries. Here we introduce a classical approach to the simulation
of general quantum circuits based on neural-network quantum states
(NQS) representations. Considering a set of universal quantum gates,
we derive rules for exactly applying single-qubit and two-qubit $Z$
rotations to NQS, whereas we provide a learning scheme to approximate
the action of Hadamard gates. Results are shown for the Hadamard and
Fourier transform of entangled initial states for systems sizes and
total circuit depths exceeding what can be currently simulated with
state-of-the-art brute-force techniques. The overall accuracy obtained
by the neural-network states based on Restricted Boltzmann machines
is satisfactory, and offers a classical route to simulating highly-entangled
circuits. In the test cases considered, we find that our classical
simulations are comparable to quantum simulations affected by an incoherent
noise level in the hardware of about $10^{-3}$ per gate.
\end{abstract}
\maketitle

\paragraph*{Introduction.- }

Quantum algorithms offer exponential speedup over the best known classical
algorithms for several interesting problems, ranging from quantum
simulation \cite{Feynman82simulatingphysics,lloyd_universal_1996}
to integer factoring \cite{shor_polynomial-time_1997}. While the
origin of the speedup varies from one quantum algorithm to another
\cite{divincenzo_quantum_1995,RevModPhys.82.1}, an essential ingredient
is the ability to efficiently modify and retrieve information stored
in the large Hilbert space in which the quantum wave function lives.
With quantum computers being actively developed \cite{neumann_multipartite_2008,mariantoni_implementing_2011,johnson_quantum_2011,blatt_quantum_2012,debnath_demonstration_2016,kandala_hardware-efficient_2017,hempel_quantum_2018},
it is becoming crucial to understand what practical applications these
platforms should target, and what problems instead are better suited
for classical algorithms. Indeed, delineating the so-called ``quantum
supremacy''\textendash the point at which a quantum computer has
solved a problem that a classical computer cannot solve in the foreseeable
future\textendash is a topic of very active current research \cite{Lund2017,Harrow2017}
that pushes the limits of classical computation \cite{Boixo2018,boixo2017,haener2017,pednault2017,chen2018,chen2018b,li2018}.
To this end, useful guidance can be obtained from a comprehensive
understanding of which interesting quantum algorithms that can be
efficiently approximated using classical computers.

At first sight, the nominal exponential complexity of a generic quantum
state suggests that efficient classical simulation of large quantum
algorithms is impossible. However, in several interesting applications
the \emph{practical} complexity of the quantum state is greatly reduced,
and heuristic classical approaches can be extremely successful at
simulating the circuit at hand. Several classical approaches exploit
the explicit structure of specific quantum circuits to provide polynomially,
or effectively polynomially scaling simulation algorithms. Noticeable
examples are circuits that are constituted of Clifford group gates
\cite{1998quant.ph..7006G,jozsa_matchgates_2008,jozsa_jordan-wigner_2015},
entanglement-limited circuits \cite{Jozsa2011,PhysRevLett.91.147902},
and circuits with restricted topological and depth properties \cite{2005quant.ph.11069M,PhysRevLett.96.170503,2006quant.ph..3163J,van_den_nest_simulating_2011}.
In addition, techniques originally developed in the context of many-body
quantum theory can be also used to efficiently simulate specific classes
of quantum circuits. Particularly successful are approaches based
on Matrix Product States (MPS) \cite{white_density_1992,fannes1992,ostlund1995},
allowing to efficiently simulate the Quantum Fourier Transform of
initial states of manageable bond dimension, typically arising in
1D physical systems \cite{2007NJPh....9..146B}. Of course these approaches
cannot work in general, and there are thus many quantum algorithms
of interest for which efficient classical algorithms are unknown.
These most chiefly include quantum algorithms involving high entanglement,
and with underlying geometries beyond the one-dimensional case amenable
to MPS simulations. In this paper, we ask how we can extend the current
abilities of classical simulation in these situations.

In recent times, machine learning techniques have been introduced
as a novel approach to simulate highly entangled quantum systems.
Key component of these approaches is a compact representation of the
many-body quantum state, based on artificial neural networks \cite{carleo_solving_2016}.
This approach has so far been used for the simulation of systems relevant
for many-body quantum theory, including spins \cite{deng_quantum_2017,glasser_neural-network_2018,choo_symmetries_2018,kaubruegger_chiral_2018,carleo_constructing_2018},
bosons \cite{saito_machine_2017,saito_method_2018,saito_solving_2017,choo_symmetries_2018},
and fermions \cite{nomura_restricted-boltzmann-machine_2017}. Having
a compact representation of the many-body state naturally leads to
applications to quantum computing. For example, neural-network representations
of quantum states have been used to learn many-qubit states from experimental
measurements. Given the success in efficiently reconstructing the
state of possibly large quantum computers \cite{torlai_neural-network_2018,rocchetto_learning_2018,torlai_latent_2018},
and highly-entangled states \cite{deng_exact_2016,deng_quantum_2017,chen_equivalence_2017},
it is natural to ask whether a classical algorithms based on artificial
neural networks and stochastic learning can be devised to simulate
quantum circuits.

In this paper, we present a general method to approximate the unitary
transformations that comprise quantum circuits using neural-network
quantum states based on complex-valued restricted Boltzmann machines
and Monte Carlo sampling. This is achieved via a stochastic framework
to learn the target quantum state after each gate in the circuit.
In particular, we demonstrate the effectiveness of the method on fundamental
primitives of quantum algorithms, such as the Hadamard and the Quantum
Fourier transforms. As non-trivial entangled test input states, we
consider the critical ground states of the transverse field Ising
model (TFIM) in both one and two dimensions.

\paragraph{Neural-network states.- }

Consider a quantum system consisting of $N$ qubits. Here we use a
representation of the many-body state associated to this system in
terms of a neural-network quantum state (NQS). More specifically,
we consider a Restricted Boltzmann machine (RBM) architecture \cite{Hinton02,hinton_reducing_2006,lecun_deep_2015}.
The RBM consists of a visible layer of $N$ nodes corresponding to
the qubit degrees of freedom, and a layer of M latent variables ($h_{1},h_{2},...,h_{M}$).
In the following, we work in the basis of eigenstates of the Pauli
$Z$ operator on each site, and specify a basis state through bitstrings
$|\mathcal{B}\rangle\equiv|B_{1},B_{2},...,B_{N}\rangle$. Here the
$B$ variables are quantum numbers that can take the values $\left\{ 0,1\right\} $,
and $Z|B\rangle=(-1)^{B}|B\rangle$. Given an input bitstring $\mathcal{B}$,
the RBM is taken to output a complex number corresponding to the unnormalized
wave-function amplitude $\langle\mathcal{B}|\Psi\rangle\equiv\Psi(\mathcal{B})$.
This network description corresponds to the following variational
expression for the quantum states \cite{carleo_solving_2016}:
\begin{multline}
\Psi_{\mathcal{W}}(\mathcal{B})=\exp\left(\sum_{j}^{N}a_{j}B_{j}\right)\times\\
\times\prod_{k}^{M}\left[1+\exp\left(b_{k}+\sum_{j}W_{jk}B_{j}\right)\right],\label{eq:psirbm}
\end{multline}
where the lower-script $\mathcal{W}$ denotes the dependence on a
set of complex variational parameters, including the visible bias
$a_{j}$, the hidden bias $\beta_{k}$ and the weights $W_{jk}$.
\begin{figure}
\includegraphics[width=1\columnwidth]{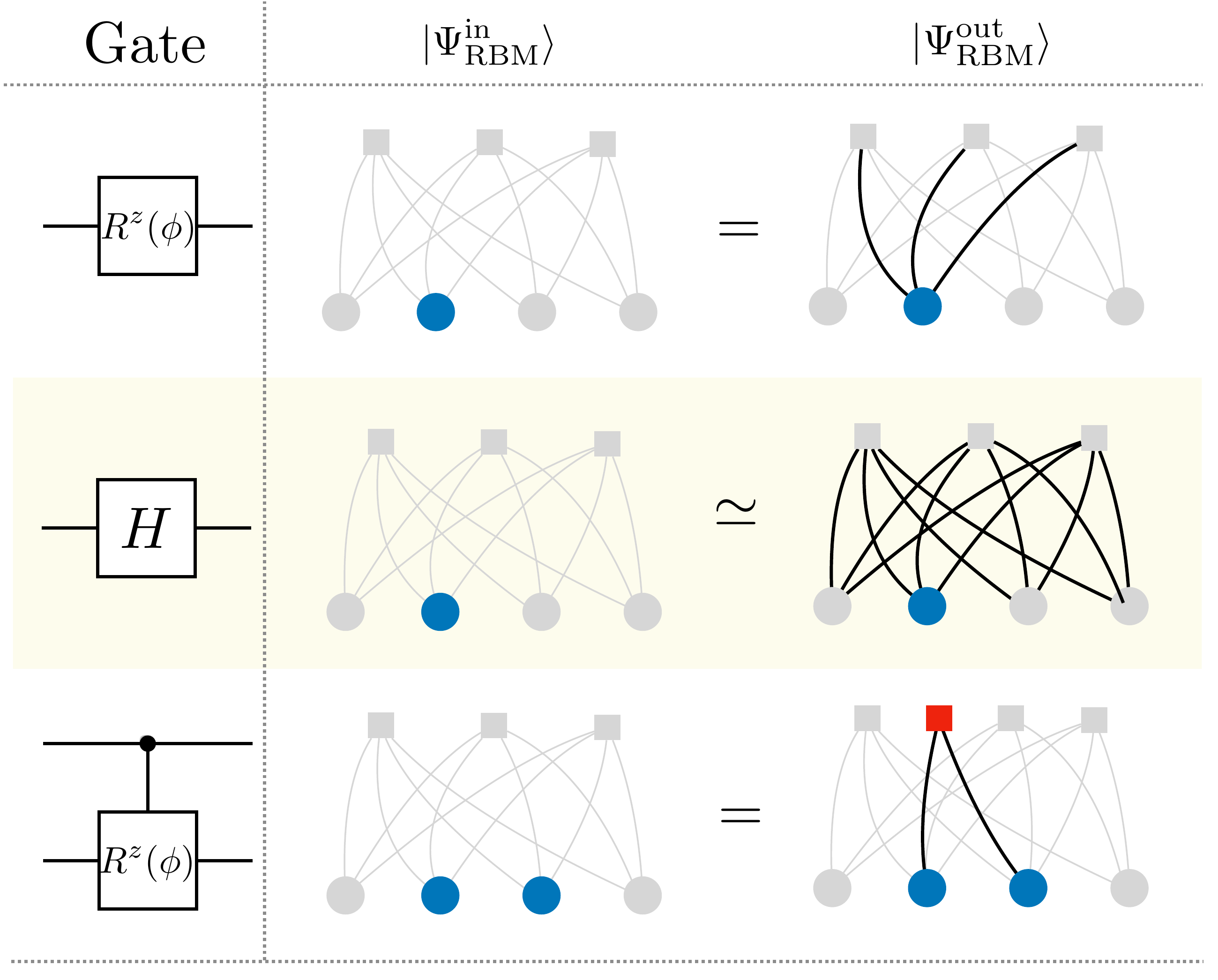}\caption{\label{fig:gates_rbm}\textbf{Action of a set of universal gate on
a restricted Boltzman machine state.} Here, $|\Psi_{\mathrm{RBM}}^{\mathrm{in}}\rangle$
represents the input state to which the gates in leftmost column are
applied, $|\Psi_{\mathrm{RBM}}^{\mathrm{out}}\rangle$ is instead
the output state. Single-qubit $Z$ rotations (upper panels) acting
on a given qubit (in blue) result in local weight modifications, $\mathrm{CR^{Z}(\phi)}$
gates (bottom panels) acting on two qubits (in blue) require instead
the introduction of an extra hidden neuron (in red). Those two families
of gates can be applied exactly, whereas the Hadamard gate (middle
panels) is approximated through the numerical scheme described in
the text. }
\end{figure}

We then consider a quantum circuit, whose action is fully specified
by a sequence of $N_{g}$ local gates $\mathcal{G}_{p}$ where $p=1,..,N_{g}$.
Any unitary operation and hence any quantum circuit can be approximated
to arbitrary accuracy in terms of a set of universal gates. In the
following, we consider the universal set of gates comprising single-qubit
rotations around the $Z$ axis ($R^{Z}(\phi)$), the Hadamard gate
($H$) and controlled rotations around the $Z$ axis ($\mathrm{CR^{Z}(\phi)}$).
In order to simulate quantum computing with a neural-network ansatz,
we then devise strategies to apply such gates to the states \eqref{eq:psirbm}.
Specifically, for a given gate $\mathcal{G}_{p}$ we look for a solution
to $\langle\mathcal{B}|\Psi_{\mathcal{W^{\prime}}}\rangle=\langle\mathcal{B}|\mathcal{G}_{p}|\Psi_{\mathcal{W}}\rangle$,
where the new set of weights $\mathcal{W^{\prime}}$ should be determined
in such a way that this equation is satisfied for all the possible
values of bistrings $\mathcal{B}$.

For the particular choice of NQS used here, all gates that are diagonal
in the computational basis can easily be applied exactly. In the case,
of single-qubit gates, these are (up to irrelevant global phases)
all given by the diagonal matrix $R^{z}(\phi)=\mathrm{Diag(1,e^{i\phi})}$.
The action of $R^{Z}(\phi)$ on a given qubit $l$ yields amplitudes
$\langle\mathcal{B}|R_{l}^{Z}(\phi)|\Psi_{\mathcal{W}}\rangle=e^{iB_{l}\phi}\Psi_{\mathcal{W}}(\mathcal{\mathcal{B}})$.
The action of the gate can be exactly reproduced with the choice $\mathcal{W}^{\prime}=\{\alpha^{\prime},\beta,W\}$,
where $a_{j}^{\prime}=a_{j}+\delta_{jl}i\phi.$ For two qubits, we
consider the controlled $Z$ rotations, and together with the single-qubit
rotations generate all diagonal gates. Their action on a given state,
$\langle\mathcal{B}|\mathrm{CR^{Z}}_{lm}(\phi)|\Psi_{\mathcal{W}}\rangle=e^{iB_{l}B_{m}\phi}\Psi_{\mathcal{W}}(\mathcal{\mathcal{B}})$,
can also be exactly satisfied, apart from a trivial global normalization
constant, by introducing an extra hidden unit $h_{[c]}$ coupled only
to qubits $l$ and $m$ through the weights $W_{l[c]}=-W_{m[lm]}=2A(\phi)$,
with $A(\phi)=\mathrm{arcosh}(e^{-i\phi/2})$ as derived in detail
in the Appendix. Also, this gate requires a change in the visible
bias, in such a way that $a_{l}^{\prime}=a_{l}+i\phi/2-A(\phi)$ and
$a_{m}^{\prime}=a_{m}+i\phi/2+A(\phi)$. Note that the maximum number
of hidden units that need to be added to implement all such two-qubit
gates is $N^{2}$. The action of circuits containing only $R^{Z}(\phi)$
and $\mathrm{CR^{Z}(\phi)}$ gates thus can be efficiently simulated
in terms of RBM states, since it induces local network modifications
which are easily determined using the rules discussed above (see Fig.
\ref{fig:gates_rbm} for a schematic representation of how the RBM
state is modified in those cases).
\begin{figure*}[t]
\includegraphics[width=1.6\columnwidth]{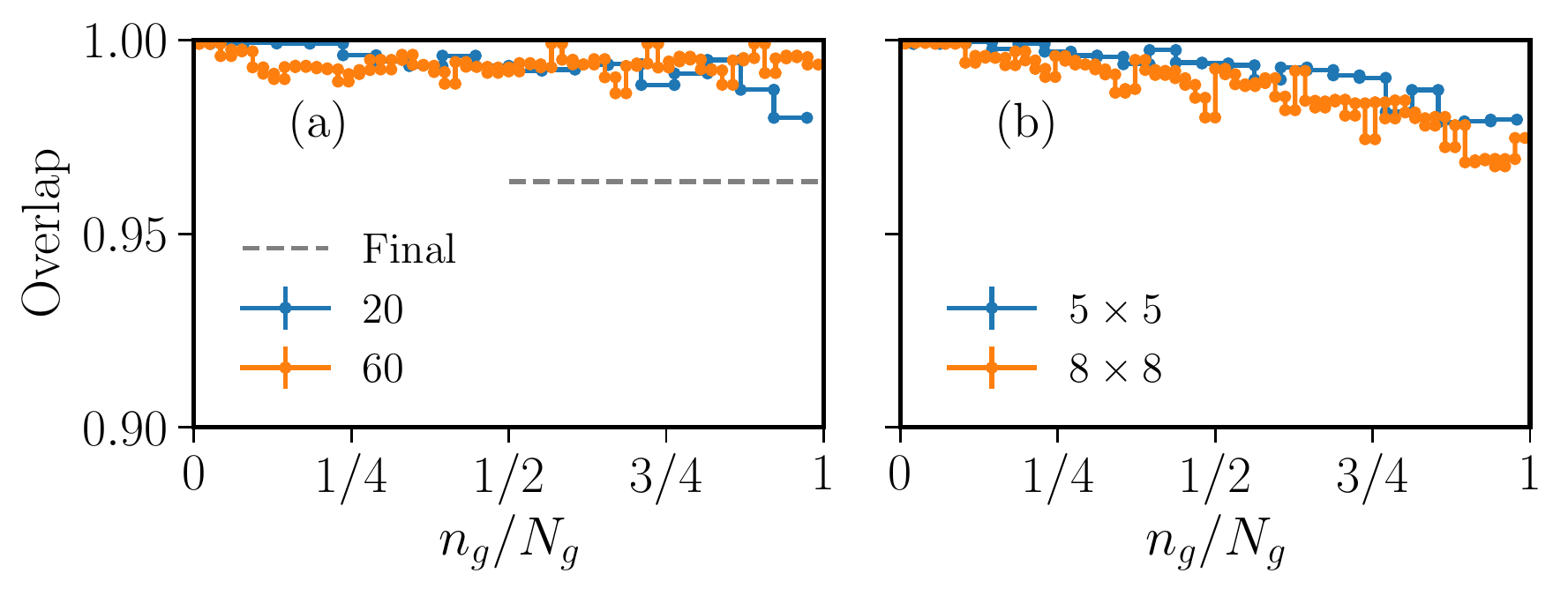}

\includegraphics[width=1.6\columnwidth]{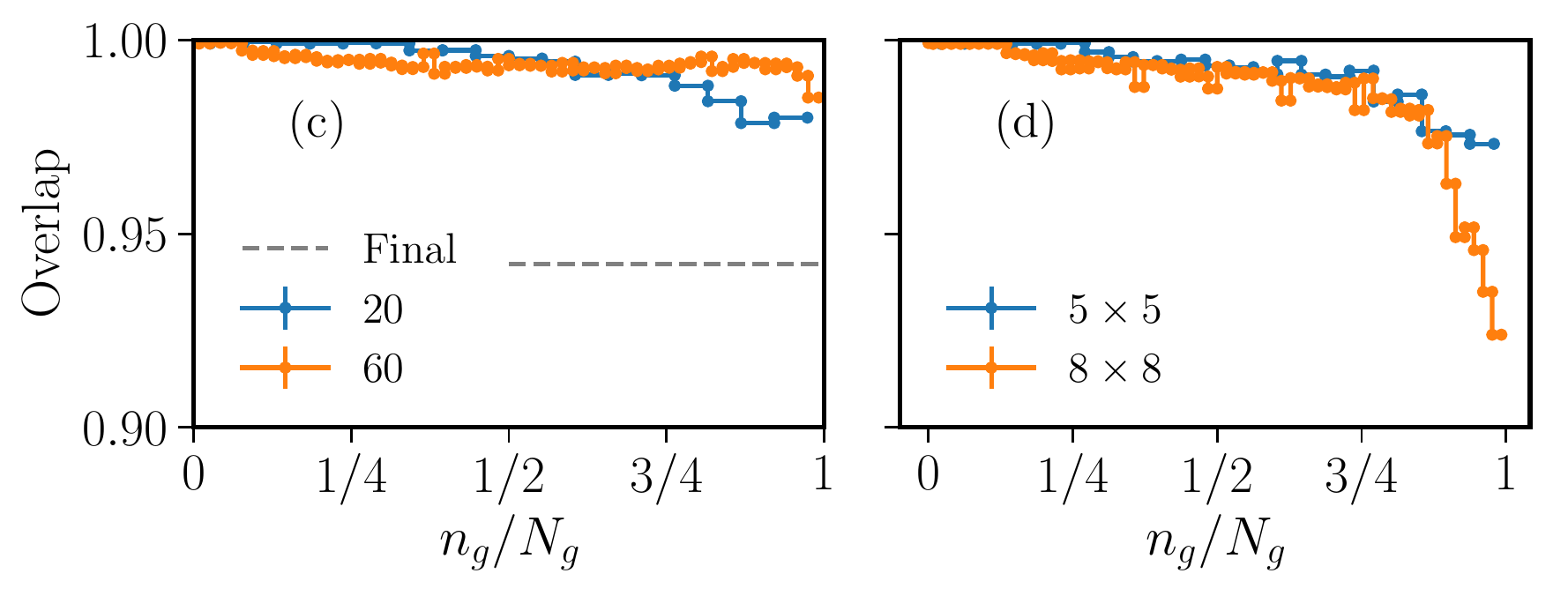}\caption{\label{fig:transforms}\textbf{Simulation of quantum transforms with
neural-network quantum states.} Hadamard transform results are shown
in (a,b), whereas Truncated Fourier Transform results are shown in
(c,d). Input for the transforms are the ground-states of the Transverse-Field
Ising model for one-dimensional chains with periodic boundaries (a,c)
and two-dimensional square lattices (b,d). Values of the transverse
field are chosen to be at or close to the critical points, thus $\Gamma/J=1$
and $\Gamma/J=3$, respectively. Dots connected by continuous lines
show the intermediate fidelities obtained when approximate the action
of the Hadamard gates in the circuit. $N_{\mathrm{g}}$ denotes here
the total number of Hadamard gates applied. The horizontal dashed
lines shown the final overlap between the NQS approximation at the
end of the circuit, and the exact state, obtained for the smaller
system size ($N=20$ for one-dimensional chains). }
\end{figure*}

\paragraph{Approximating Hadamard gates.-}

In order to provide a complete scheme for the classical simulation
of quantum circuits with NQS, we further need to specify the action
of the Hadamard gate. In contrast to the previously discussed unitaries,
in the general case it is hard to find exact strategies to efficiently
apply the Hadamard gate to an NQS. The exact application of the Hadamard
gate results in the introduction of an additional layer in the Boltzmann
machine, thus going beyond the NQS form \cite{gao_efficient_2017}.
As a consequence of the additional deep layer, the sampling from the
evolved states quickly becomes intractable as the circuit depth grows.
Here, we instead consider an approximate strategy, relying on a generalized
variational treatment of the quantum circuit in the pure RBM form.
Given an initial variational state $|\Psi_{\mathcal{W}}\rangle$,
our goal is to devise an efficient numerical scheme to obtain an optimal
representation of the many-body quantum state after the Hadamard gate,
$|\Phi\rangle=H|\Psi_{\mathcal{W}}\rangle$, such that $|\Psi_{\mathcal{W}^{\prime}}\rangle\simeq|\Phi\rangle$
for a set of parameters $\mathcal{W}^{\prime}$ to be determined.
Specifically, we consider the negative log-overlap:
\begin{equation}
L(\Psi_{\mathcal{W}^{\prime}},\Phi)=-\log\left[\frac{|\langle\Psi_{\mathcal{W^{\prime}}}|\Phi\rangle|}{||\Psi_{\mathcal{W}^{\prime}}\rangle|||\Phi\rangle|}\right],\label{eq:loginf}
\end{equation}
which attains a minimum $L=0$ when the two states are equal, and
devise a procedure to minimize it with respect to $\mathcal{W^{\prime}}$.
In the language of machine learning, this corresponds to the loss
function. In analogy with the time-dependent variational Monte Carlo
scheme \cite{carleo_localization_2012,carleo_light-cone_2014}, this
quantity can be minimized using a stochastic procedure. In particular,
the gradient with respect to the $k$-th variational parameter, $p_{k}$,
can expressed in terms of expectation values:
\begin{eqnarray}
\partial_{p_{k}}L(\Psi_{\mathcal{W}^{\prime}},\Phi) & = & \langle\mathcal{O}_{k}^{\star}(\mathcal{B})\rangle_{\Psi}-\frac{\left\langle \frac{\Phi(\mathcal{B})}{\Psi(\mathcal{B})}\mathcal{O}_{k}^{\star}(\mathcal{B})\right\rangle _{\Psi}}{\left\langle \frac{\Phi(\mathcal{B})}{\Psi(\mathcal{B})}\right\rangle _{\Psi}},\label{eq:gradlog}
\end{eqnarray}
where we have introduced the variational operators $\mathcal{O}_{k}(\mathcal{\mathcal{B}})=\partial_{p_{k}}\log\Psi_{\mathcal{W}^{\prime}}(\mathcal{B})$,
and $\langle F(\mathcal{B})\rangle_{\Psi}\equiv\sum_{\mathcal{B}}F(\mathcal{B})|\Psi(\mathcal{B})|^{2}/\sum_{\mathcal{B}}|\Psi(\mathcal{B})|^{2}$
denote expectation values over the variational state $\Psi_{\mathcal{W^{\prime}}}$.
The minimization of the negative log-overlap with respect to the parameters
of the RBM is then achieved through an iterative scheme where at each
iteration stochastic estimates of the gradient are obtained through
Eq. \eqref{eq:gradlog}. The network parameters are then updated with
a Stochastic Gradient Descent optimization method \cite{kingma_adam:_2014}.
This stochastic approach can then be used to systematically optimize
the log-overlap even on large systems, inaccessible to exact simulation
approaches. Details concerning the sampling procedure, and the optimization
steps are discussed in the Appendix.
\begin{figure*}[t]
\includegraphics[width=1.6\columnwidth]{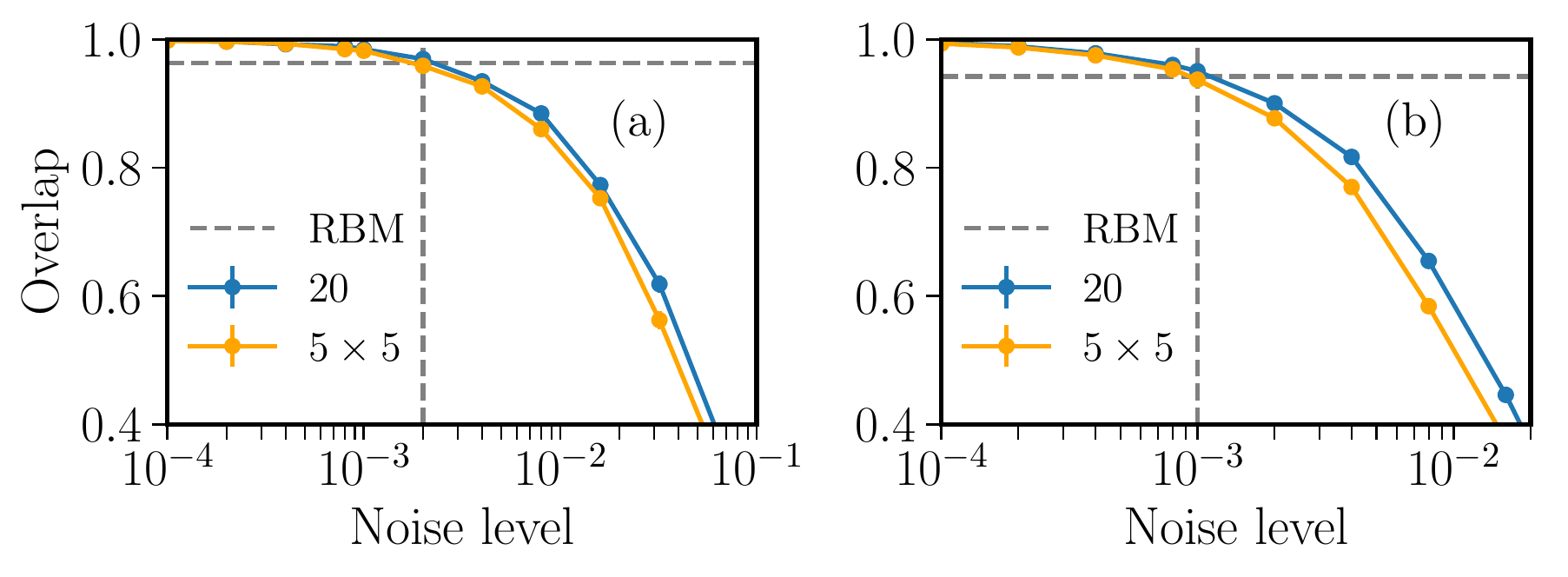}\caption{\label{fig:noise}\textbf{Comparing the effect of hardware noise to
the variational error.} Continuous lines show the behavior of the
overlap as a function of the noise level in a minimal model consisting
of depolarizing channels after each gate. The noise model is applied
only to the quantum transforms, whereas the preparation of the initial
state is assumed to be exact in the hardware. Results are shown for
both the Hadamard (a) and the truncated Fourier (b) transforms. Horizontal
dashed lines correspond to the NQS final overlap for the smaller 1d
systems. }
\end{figure*}

\paragraph{Hadamard and Fourier transform.-}

To validate the effectiveness of this scheme in approximating prototypical
quantum circuits, we benchmark our approach on the simulation of quantum
transforms of entangled initial states. Specifically, the initial
states are prepared considering the transverse-field Ising Hamiltonian:
\begin{eqnarray}
\mathcal{H} & = & -\Gamma\sum_{i}X_{i}+J\sum_{\langle i,j\rangle}Z_{i}Z_{j},
\end{eqnarray}
where the interaction terms runs over pairs of nearest neighbors on
a lattice. These states exemplify quantum states commonly encountered
in quantum simulation. We consider both one-dimensional chains and
two-dimensional square lattices with periodic boundary conditions,
and prepare an initial NQS in the ground-state of $\mathcal{H}$ at
or close to the critical point ($\Gamma/J=1$ in 1d, and $\Gamma/J=3$
for the 2d square lattice). We prepare NQS variational ground states
with $\alpha=1$, which in previous work was already shown to yield
an accurate description of the ground-state, $|\Phi_{0}\rangle$.

The first transform we consider is the Hadamard transform, in which
the output state is given by $|\Phi_{HT}\rangle=H_{1}\dots H_{N}|\Phi_{0}\rangle$,
i.e. the corresponding quantum circuits contains Hadamard gates on
each qubit. We further apply our approach to the quantum Fourier transform,
one of the fundamental building blocks for the most important quantum
algorithms, including Shor \cite{shor_polynomial-time_1997}, phase
estimation \cite{kitaev_quantum_1995}, and more \cite{Nielsen:2011:QCQ:1972505}.
For simplicity, we have considered here the truncated Fourier transform,
where the final state is given by $|\Phi_{\mathrm{TFT}}\rangle=H_{N}\dots B_{2,4}A_{2,3}H_{2}B_{1,3}A_{1,2}H_{1}|\Phi_{0}\rangle,$where
the controlled gates are $A=\mathrm{CR^{Z}}(\pi/2)$ and $B=\mathrm{CR^{Z}}(\pi/4)$.

Numerical results for the Hadamard transform are shown in Fig. \ref{fig:transforms}
(a,b), and in (c,d) for the truncated Fourier transform. Results for
both one- and two-dimensional initial states are presented. The intermediate
fidelity, as obtained stochastically minimizing \eqref{eq:loginf}
at each step involving an Hadamard gate is shown as a function of
the total circuit depth. For both one and two-dimensional circuits
we find satisfactorily high intermediate overlaps ($>0.96$) with
RBM states of fixed number of hidden units, $\alpha=1$. To the best
of our knowledge, the circuits shown here for the largest systems
cannot be exactly simulated with existing state-of-the-art brute-force
approaches.

In order to assess the overall quality of the variational approximation,
beyond the individual gates fidelity, for small one-dimensional systems
we compute the overlap between the exact output state and the approximate
output variational state (horizontal dashed lines). As expected, the
overall fidelity is lower than the intermediate ones, yet not significantly
lower than what found for the intermediate steps. Note in particular
that the fidelity at the end is better than the product of the intermediate
fidelities over the whole evolution, suggesting that there is a cancellation
of errors in the final state.

\paragraph{Noisy quantum computing.-}

One of the major challenges for quantum computers is controlling and
mitigating the effect of noise caused by interactions of the quantum
system with the environment. As a result of such decoherence, the
output of a quantum computer is affected by a loss of fidelity with
the exact target state, if error correction is not taken into account.
It is therefore interesting to compare the variational error in the
classical NQS simulation to the error due to a noisy quantum hardware.
To simulate the effect of noise, we consider here a simple Pauli noise
channel, which approximates the depolarizing channel \cite{knill_randomized_2008,barends_superconducting_2014,barends_digitized_2016}.
This is implemented by applying, after each one qubit gate, with probability
$r$one of the three randomly selected Pauli operators and with probability
$1-r$ the qubit is left untouched. For the two-qubit gates one of
the 15 combinations of Pauli operators, where the identity on one
of the qubits is included, is applied with probability $r$ and the
qubit pair is left untouched with probability $1-r$. Here the parameter
r has the role of a noise level. Results are shown in Fig. \ref{fig:noise}
for both the Hadamard transform (a) and the truncated Fourier transform
(b). We find similar qualitative and quantitative behavior of the
final overlap versus the noise level for both one and two-dimensional
initial states. The final overlap achieved with the RBM variational
ansatz in 1d is shown as horizontal dashed lines, and correspond to
an effective noise level of about $10^{-3}$. Notice that the noise
model considered here takes into account hardware error only for the
quantum transforms circuits, and not for the circuits necessary to
prepare the initial states. The effective noise level for the complete
circuits is expected to be significantly lower than $10^{-3}$.

\paragraph{Discussion.-}

We have introduced a stochastic algorithm to perform classical simulation
of large, entangled quantum circuits. Our results show that neural-network
quantum states can approximate, with good fidelity, quantum circuits
beyond the current state of the art of brute-force classical simulation.
Our approach, while not expected to be efficient in cases where sampling
from the wave-function is practically hard, effectively enlarge the
space of interesting quantum algorithms that can be classically simulated.
Apart from the practical applications of our algorithm, several conceptual
points will hopefully be stimulated by our results. For example, the
theoretical analysis of the complexity of quantum algorithms might
find new insights coming from highly-entangled classical variational
representations of the quantum states. Finally, the results presented
here can serve as a guide for the undergoing development of quantum
hardware, setting reference values for the maximum noise levels necessary
to outperform classical approximation schemes.
\begin{acknowledgments}
We acknowledge discussions with S. Bravyi, X. Gao, D. Gosset, A. Mezzacapo,
A. Rocchetto, S. Severini, and S. Strelchuk. Neural-network quantum
states simulations are based on NetKet \cite{NetKet}.
\end{acknowledgments}

\appendix

\section{Exact Application of Quantum Gates to Restricted Boltzmann Machines}

In this Appendix we discuss a number of quantum gates whose application
to an RBM state can be performed efficiently. These include all the
single-qubit Pauli gates and Z rotations, as well as two-qubit controlled
Z rotations.

\subsection{Single-Qubit Z rotations}

The action of the single-qubit Z rotations of angle $\phi$ is fully
determined by the $2\times2$ unitary matrix
\begin{eqnarray*}
R^{Z}(\phi) & = & \left(\begin{array}{cc}
1 & 0\\
0 & e^{i\phi}
\end{array}\right)
\end{eqnarray*}
 Its action on a given qubit $l$ yields $\langle\mathcal{B}|R_{l}^{z}(\phi)|\Psi_{\mathcal{W}}\rangle=e^{i\phi B_{l}}\Psi_{\mathcal{W}}(\mathcal{\mathcal{B}})$.
Considering a RBM machine with weights $\mathcal{W}^{\prime}=\{\alpha,\beta,W\}$,
the action of the $R^{Z}(\phi)$ gate is exactly reproduced if we
satisfy $e^{B_{l}a_{l}}e^{i\phi B_{l}}=e^{B_{l}a_{l}^{\prime}},$
which has the simple solution:
\begin{eqnarray}
a_{j}^{\prime} & = & a_{j}+\delta_{jl}i\phi.
\end{eqnarray}
The action of this gate then simply modifies the local visible bias
of the RBM.

\subsection{Controlled Z rotations }

The action of a controlled Z rotations acting on two given qubits
$l$ and $m$ is determined by the $4\times4$ unitary matrix:
\begin{eqnarray}
\mathrm{CR^{Z}}(\phi) & = & \left(\begin{array}{cccc}
1 & 0 & 0 & 0\\
0 & 1 & 0 & 0\\
0 & 0 & 1 & 0\\
0 & 0 & 0 & e^{i\phi}
\end{array}\right),
\end{eqnarray}
where $\phi$ is a given rotation angle. This gate is diagonal, and
we can compactly write it as an effective two-body interaction:
\begin{eqnarray}
\left\langle \mathcal{B}\right|\mathrm{CZ}(\phi)\left|\Psi_{\mathcal{W}}\right\rangle  & = & e^{i\phi B_{l}B_{m}}\Psi_{\mathcal{W}}(Z_{1}\dots Z_{N}).
\end{eqnarray}
Since in the RBM architecture there is no direct interaction between
visible spins, this CZ interaction can be mediated through the insertion
of a dedicated extra hidden unit $h_{c}$, which is coupled only to
the qubits $l$ and $m$:
\begin{eqnarray}
\left\langle \mathcal{B}\right|\mathrm{CZ}(\phi)\left|\Psi_{\mathcal{W}}\right\rangle  & = & e^{\Delta a_{l}B_{l}+\Delta a_{m}Z_{m}}\sum_{h_{c}}e^{W_{lc}B_{l}h_{c}+W_{mc}B_{m}h_{c}}\\
 & = & e^{\Delta a_{l}B_{l}+\Delta a_{m}B_{m}}\times\nonumber \\
 &  & \times(1+e^{W_{lc}B_{l}+W_{mc}B_{m}})\Psi_{\mathcal{W}}(\mathcal{B}),
\end{eqnarray}
where the new weights $W_{lc}$ and $W_{mc}$ , and visible units
biases $a_{l}^{\prime}=a_{l}+\Delta a_{l}$, $a_{m}^{\prime}=a_{m}+\Delta a_{m}$
are determined by the equation:
\begin{eqnarray}
e^{\Delta a_{l}B_{l}+\Delta a_{m}B_{m}}(1+e^{W_{lc}B_{l}+W_{mc}B_{m}}) & = & C\times e^{i\phi B_{l}B_{m}},
\end{eqnarray}
for all the 4 possible values of the qubits values $B_{l},B_{m}=\{0,1\}$
and where $C$ is an arbitrary (finite) normalization. A possible
solution for this system is:
\begin{eqnarray}
W_{lc} & = & -2\mathcal{A}(\phi)\\
W_{mc} & = & 2\mathcal{A}(\phi)\\
\Delta a_{l} & = & i\frac{\phi}{2}+\mathcal{A}(\phi)\\
\Delta a_{m} & = & i\frac{\phi}{2}-\mathcal{A}(\phi),
\end{eqnarray}
where $\mathcal{A}(\phi)=\mathrm{arccosh}\left[e^{-i\frac{\phi}{2}}\right]$.

\subsection{Pauli X gate }

We then consider an $X$ gate, acting on some given qubit $l$. In
this case, the gate just flips the qubit, and the RBM amplitudes are:
\begin{eqnarray*}
\langle\mathcal{B}|X_{l}|\Psi_{\mathcal{W}}\rangle & = & \langle B_{1}\dots\bar{B}_{l}\dots B_{N}|\Psi_{\mathcal{W}}\rangle,
\end{eqnarray*}
therefore since $\bar{B}_{l}=(1-B_{l})$, we must satisfy
\begin{eqnarray}
(1-B_{l})W_{lk}+b_{k} & = & B_{l}W_{lk}^{\prime}+b_{k}^{\prime},\\
(1-B_{l})a_{l} & = & B_{l}a_{l}^{\prime}+C,
\end{eqnarray}
for all the (two) possible values of $B_{l}=0,1$. The solution is
simply:
\begin{eqnarray}
W_{lk}^{\prime} & = & -W_{lk}\\
b_{k}^{\prime} & = & b_{k}+W_{lk}\\
a_{l}^{\prime} & = & -a_{l}\\
C & = & a_{l}.
\end{eqnarray}
whereas all the $a_{j}$ and the other weights $W_{jk}$ with $j\neq l$
are unchanged.

\subsection{Pauli Y gate}

A similar solution is found also for the Y gate, with the noticeable
addition of extra phases with respect to the $X$ gate:

\begin{eqnarray}
W_{lk}^{\prime} & = & -W_{lk}\\
b_{k}^{\prime} & = & b_{k}+W_{lk}\\
a_{l}^{\prime} & = & -a_{l}+i\pi\\
C & = & a_{l}+\frac{i\pi}{2}.
\end{eqnarray}
whereas all the $a_{j}$ and the other weights $W_{jk}$ with $j\neq l$
are unchanged.

\subsection{Pauli Z gate}

For a $Z$ gate acting on qubit $l$, we have:
\begin{eqnarray}
\langle\mathcal{B}|Z_{l}|\Psi_{\mathcal{W}}\rangle & = & (-1)^{B_{l}}\langle\mathcal{B}|\Psi_{\mathcal{W}}\rangle,
\end{eqnarray}
therefore we must satisfy $e^{B_{l}a_{l}}(-1)^{B_{l}}=e^{B_{l}a_{l}^{\prime}},$
which has the simple solution:
\begin{eqnarray}
a_{l}^{\prime} & = & a_{l}+i\pi,
\end{eqnarray}
whereas all the other weights and biases are unchanged.

\section{Stochastic Approximation of the Hadamard Gate}

The key numerical component of our approach is the stochastic framework
needed to find network parameters $\mathcal{W}^{\prime}$ such that
$|\Psi_{\mathcal{W}^{\prime}}\rangle\simeq|\Phi\rangle$ , where $|\Phi\rangle=H_{l}|\Psi_{\mathcal{W}}\rangle$
is the exact state after a Hadamard gate has been applied on some
qubit $l$. The overlap between the two states is the key quantity
to measure how close these two states are, and reads
\begin{eqnarray}
\mathrm{O}(\mathcal{W}^{\prime}) & = & \sqrt{\frac{|\langle\Psi_{\mathcal{W}^{\prime}}|\Phi\rangle|^{2}}{\langle\Psi_{\mathcal{W}^{\prime}}|\Psi_{\mathcal{W}^{\prime}}\rangle\langle\Phi|\Phi\rangle}}\nonumber \\
 & = & \sqrt{\left\langle \frac{\Phi(\mathcal{B})}{\Psi_{\mathcal{W}^{\prime}}(\mathcal{B})}\right\rangle _{\Psi}\left\langle \frac{\Psi_{\mathcal{W}^{\prime}}(\mathcal{B})}{\Phi(\mathcal{B})}\right\rangle _{\Phi}^{\star}},\label{eq:overlapstoch}
\end{eqnarray}
where $\langle F(\mathcal{B})\rangle_{A}\equiv\sum_{\mathcal{B}}F(\mathcal{B})|A(\mathcal{B})|^{2}/\sum_{\mathcal{B}}|A(\mathcal{B})|^{2}$
denote expectation values over the variational state ($A(\mathcal{B})=\Psi_{\mathcal{W^{\prime}}}(\mathcal{B})$)
and over the exact state $(A(\mathcal{B})=\Phi(\mathcal{B}))$, respectively.
In order to stochastically compute the overlap, we then generate two
independent set of samples, one distributed according to $|\Psi_{\mathcal{W^{\prime}}}(\mathcal{B})|^{2}$
and another one according to $|\Phi(\mathcal{B})|^{2}$. These can
be obtained using standard Markov Chain Monte Carlo techniques \cite{becca_quantum_2017},
and look-up tables approaches as discussed in Ref. \cite{carleo_solving_2016}
are used to reduce the overall computational complexity of the sampling.

In addition to a stochastic expression for the overlap, it is also
possible to find a suitable stochastic expression for its derivatives
with respect to a generic network-parameter $k$. In practice, it
is more convenient to define as a loss function to be minimized:
\begin{eqnarray}
L(\mathcal{W}^{\prime}) & = & -\log\mathrm{O}(\mathcal{W}^{\prime}),\label{eq:logoverlap}
\end{eqnarray}
which attains a minimum $(L=0$) when the two states are equal. In
this case, the gradients have a more compact form than for the bare
overlap, and read:
\begin{eqnarray}
\partial_{p_{k}}L(\Psi_{\mathcal{W}^{\prime}}) & = & \langle\mathcal{O}_{k}^{\star}(\mathcal{B})\rangle_{\Psi}-\frac{\left\langle \frac{\Phi(\mathcal{B})}{\Psi(\mathcal{B})}\mathcal{O}_{k}^{\star}(\mathcal{B})\right\rangle _{\Psi}}{\left\langle \frac{\Phi(\mathcal{B})}{\Psi(\mathcal{B})}\right\rangle _{\Psi}},\label{eq:gradlog-1}
\end{eqnarray}
where $\mathcal{O}_{k}(\mathcal{\mathcal{B}})=\partial_{p_{k}}\log\Psi_{\mathcal{W}^{\prime}}(\mathcal{B})$
are the variational derivatives, and estimates of the gradient can
be obtained sampling according to $|\Psi_{\mathcal{W^{\prime}}}(\mathcal{B})|^{2}$.

Once the loss function and its gradient are defined, we can use any
standard Stochastic Gradient Descent method to carry on the optimization.
In our results, we have used AdaMax \cite{kingma_adam:_2014}, and
typically initialized the parameters in a way that $\mathcal{W^{\prime}\simeq\mathcal{W}}$,
adding some small random noise to $\mathcal{W}$.

\end{document}